\documentclass[fleqn,twoside]{article}
\usepackage{espcrc2,epsfig}

\usepackage{graphicx}

\newcommand{\AmS}{{\protect\the\textfont2
  A\kern-.1667em\lower.5ex\hbox{M}\kern-.125emS}}
\title{Direct detection of WIMPs with conventional (non-cryogenic) detectors.
Experimental review
\thanks{Invited Review Talk given at the XXX
International Meeting on Fundamental Physics, IMFP2002, February
2002, Jaca, Spain.} }
\author{Angel Morales\thanks{amorales@posta.unizar.es}\\ Laboratory of Nuclear and High Energy Physics and Canfranc Underground Laboratory (LSC) \\
        University of Zaragoza \\ 50009 Zaragoza. Spain}

\begin{document}

\begin{abstract}
An overview of the current status of WIMP direct searches with
conventional detectors is presented, emphasizing strategies,
achievements and prospects.
\end{abstract}
\maketitle
\section{Introduction}

Experimental observations and robust theoretical arguments have
established that our universe is essentially non-visible, the
luminous matter scarcely accounting for one per cent of the energy
density of a flat universe. The distribution of a flat universe
($\Omega=\Omega_{M}+\Omega_{\Lambda}=1$) attributes to the dark
energy about $\Omega_{\Lambda}\sim70\%$, whereas the matter
density takes the remaining $\Omega_{M}\sim30\%$, consisting of,
both, visible ($\Omega_{l}\sim0.5\%-1\%$) and non-visible (dark)
matter. This dark component is formed by ordinary baryonic matter
($\Omega_{B}\sim 4-5\%$), (possibly made by MACHOs, jupiters,
dust, black holes, etc.) and a large fraction (up to
$\Omega_{NB}\sim25\%$) of non baryonic dark matter, supposedly
made by non-conventional, exotic particles.

These non-baryonic particles (usually named particle dark matter)
would be filling the galactic halos, at least partially, according
to a variety of models. It is supposed to be a suitable mixture
(cold and hot dark matter) to properly generate the cosmic
structures. The minimal requirements to be fulfilled by the
non-baryonic dark particles are to provide the right relic
abundance, to have non-zero mass, zero electric charge and very
weak interaction with ordinary matter.

There are several candidates to such species of matter provided by
schemes beyond the Standard Model of Particle Physics. Remarkable
examples are the axions, WIMPs (Weak Interacting Massive
PArticles) and neutrinos. Axions are pseudoscalar Nambu-Goldstone
bosons arising from the spontaneous symmetry breaking of the
Peccei-Quin $U(1)_{PQ}$ symmetry invented to solve the strong CP
problem. They are very weakly coupled to ordinary matter; examples
are the galactic DM axions of the models KSUZ and DFSZ. Axions
with large $f_{a} (\sim 10^{12}GeV)$ are good candidates to dark
matter, where $f_{a}$ is the energy scale of the PQ symmetry
breaking. The more favorable mass window for this candidate is
$10^{-5(-6)}eV<m_{a}<10^{-2(-3)}eV$.

Another popular candidate are the weak interacting, neutral and
massive particle, called WIMPs. A particularly attractive kind of
WIMPs are provided by the SUSY models, like the neutralinos (the
lightest stable particles -LSP-) of super symmetric theories. The
lowest-mass neutralino is a linear superposition of photino, zino,
higgsino, expressed as
 $\chi \equiv a_{1}\widetilde{\gamma}+a_{2}\widetilde{z}+a_{3}{\widetilde{H}}_{1}^{0}+a_{4}{\widetilde{H}}_{2}^{0}$
 where $P\equiv a_{1}^{2}+a_{2}^{2}$ and $P>0.9$ gaugino;  $P<0.1$ higgsino.
The mass window of this candidate is $GeV\leq m_{\chi}\leq TeV$.
An interesting mass region, for reasons which will become clear
later on, is that of $40 GeV\leq m_{\chi}\leq 200 GeV$.

The last candidate is the (non-zero mass) neutrino of theories
beyond the Standard Model. The neutrinos are the only candidate
known to exist, have well-known weak interaction and only a small
amount of them is needed to explain cosmic data. Their mass-window
is very wide according to the particular model (to fit also other
phenomenology of $\nu$-physics). It is to be noticed that the
masses of the non baryonic particle candidates extend along more
than 18 orders of magnitude: ~$10^{-6}$~eV - $10^{12}$~eV. On the
other hand, all the candidates have a common feature: their small
interaction cross section with ordinary matter. In the case of
WIMPs, for instance, various implementations of the Minimal SUSY
extension of the Standard Model, MSSM, lead to an interaction
cross-sections WIMP-matter, ${(\sigma_{\chi,N})}_{Th}$,
encompassing several orders of magnitude from $10^{-5}$ down to
$10^{-10}$~pb. This cross-section plays a leading role in the
estimation of the sensitivity required in particle dark matter
searches.

This talk will be devoted only to one of these candidates: the
WIMPs. Without entering into considerations about how large the
baryonic dark component of the galactic halo could be, we take for
granted that there is enough room for WIMPs in our halo to try to
detect them, either directly or through their by-products.
Discovering this form of Dark Matter is one of the big challenges
in Cosmology, Astrophysics and Particle Physics.

WIMPs can be looked for either directly or indirectly through its
byproducts. The indirect detection of WIMPs proceeds currently
through two main experimental lines: either by looking in cosmic
rays experiments for positrons, antiprotons, or other antinuclei
produced by the WIMPs annihilation in the halo (like in the
CAPRICE, BESS, AMS, GLAST, VERITAS, MAGIC experiments ...), or by
searching in large underground detectors (SUPERKAMIOKANDE, SNO,
SOUDAN, MACRO) or underwater neutrino telescopes (BAIKAL, AMANDA,
ANTARES, NESTOR) for upward-going muons produced by the energetic
neutrinos emerging as final products of the WIMPs annihilation in
celestial bodies (Sun, Earth...).

The direct detection of WIMPs relies in measuring the nuclear
recoil produced by their elastic scattering off target nuclei in
suitable detectors. The signal rate depends of the type of WIMP
and interaction, whereas simple kinematics says that the energy
delivered in the WIMP-nucleus interaction is very small. In the
case of WIMPs of $m\sim GeV~\mathrm{to}~\textit{TeV}$ and
$v\sim10^{-3}c$ the nuclear recoil energy in the laboratory frame
$E_{R}=\frac{\mu^{2}}{M}v^{2}(1-cos\theta)$ is in the range from 1
to 100 keV. $M$ is the nuclear mass, $\mu$ the ($m, M$) reduced
mass and $\theta$ the WIMP-nucleus (c. of m.) scattering angle.
Even small the recoil energy, only a fraction $QE_{R}=E_{vis}
(\equiv E_{eee})$ of it is visible in the detector, depending on
the type of detector and target and on the mechanism of energy
deposition. The so-called Quenching Factor Q is essentially unit
in thermal detectors whereas for the nuclei used in conventional
detectors it ranges from about 0.1 to 0.6. For instance for a Ge
nucleus only about 1/4 of the recoil energy goes to ionization.

On the other hand, the smallness of the neutralino-matter
interaction cross-section makes the rates of the nuclear recoil
looked for very small. The variety of models and parameters used
to describe the Astrophysics, Particle Physics and Nuclear Physics
aspects of the process makes, as stated before, the
neutralino-nucleus interaction rate encompass several orders of
magnitude, going from 10 to $10^{-5}$ c/kg day, according to the
SUSY model and parameters \cite{Bottino}. In fact, it is not
higher than $10^{-2}$ c/kg day, for the neutralino parameters
which provide the most favourable relic density.

The rare ($\leq 10^{-2}$ c/kg.day) and small (keV range) WIMP
signal falls in the low energy region of the spectrum, where the
radioactive and environmental background accumulate at much faster
rate and with similar spectral shape. That makes WIMP signal and
background practically undistinguishable. In conclusion, due to
the properties of the expected signals, the direct search for
particle dark matter through their scattering by nuclear targets
requires ultralow background detectors of a very low energy
threshold, endowed, when possible, with background discrimination
mechanisms. All these features together make the WIMP detection a
formidable experimental challenge.

This review will deal with the efforts currently being done in the
direct search of WIMPs illustrated by a few experiments. Only
conventional, non-cryogenic detectors, will be considered in this
review (the case for cryogenics detectors will be addressed in the
review of L. Mosca, in these Proceedings). The signals to be
expected, their main features, the techniques employed and the
achievements accomplished will be overviewed. The current results
and prospects will be also sketched.

\section{Detecting WIMPs}

The method to explore wether there exists or not a WIMP signal
contribution in the experimental data is rather simple: one
compares the predicted event rate with the observed spectrum; if
the former turns out to be larger than the measured one, the
particle which would produce such event rate can be ruled out as a
Dark Matter candidate. That is expressed as a contour line
$\sigma$(m) in the plane of the WIMP-nucleon elastic scattering
cross section versus the WIMP mass. For each mass m, those
particles with a cross-section above the contour line $\sigma$(m)
are excluded as dark matter. The level of background sets,
consequently, the sensitivity of the experiment to eliminate
candidates or in constraining their masses and cross sections.
Thus, the first effort to be done in looking for WIMPs must be on
diminishing the general background and then in discriminating the
background from the signal (nucleus recoil generated by WIMPs).

However, this mere comparison of the expected signal with the
experimentally observed spectrum it is not expected, in principle,
to lead to the WIMP's detection, unless one reach a reliable
zero-background ideal spectrum. In the general case, the real
spectrum, even small, which has typically the same shape than that
of the signal, could be due to pure background sources. After the
identification and rejection of most of background sources, there
still exists a worrisome background originated by the neutrons
which produce also nuclear recoils similar to the produced by
WIMPs.

This ultimate background must be identified and eliminated.
However, a convincing proof of the detection of WIMPs would need
to invent and discover signatures in the data characteristic of
the WIMPs but not of the background. There exist temporal and
spatial asymmetries specific of the WIMP interaction, which are
difficult to be faked by the background or by instrumental
artifacts. They are due to the kinematics of the motion of the
Earth (and of our detectors) in the galactic halo. These
signatures are an annual modulation of the rate and a directional
asymmetry of the nuclear recoil (see later on). Both types of
asymmetries are not characteristics of the background and, in
principle, can be used as identification labels of the WIMPs. The
only distinctive signature seriously investigated up to now is the
annual modulation of the WIMP signal rate due to the seasonal
variation produced by the Earth's motion with respect to the Sun.
Several experiments have looked for such seasonal variations of
the rates and, in fact, the DAMA experiment, after four yearly
period of data has found an annual modulation at the $3\sigma$
level, which has been associated, by the DAMA collaboration, to
the existence of a WIMP.

From the experimentalist point of view, to detect WIMPs one need
first to narrow the window ($\sigma$,m) of its possible existence,
by means of the exclusion plots, and then try its identification
through one, or more, of its distinctive features.

The detectors used so far in the quest for WIMPs (and references
later on) are: ionization detectors of Ge (IGEX, COSME, H/M, HDMS)
and of Si (UCSB), scintillation crystals of NaI (ZARAGOZA, DAMA,
UKDMC, SACLAY, ELEGANTS), liquid or liquid-gas Xenon detectors
(DAMA, UCLA, UKDMC, ZEPLIN), calcium fluoride scintillators
(MILAN, OSAKA, ROMA), thermal detectors (bolometers) with sapphire
absorbers (CRESST, ROSEBUD), with tellurite absorbers, (MIBETA,
CUORICINO) or with germanium absorbers (ROSEBUD) as well as
bolometers which also measure the ionization, like that of Si
(CDMS) and of Ge (CDMS, EDELWEISS). New techniques have been
recently incorporated. Worth to be mentioned are: scintillating
bolometers of calcium tungstate which measure heat and high
(CRESST and ROSEBUD), and of BGO (ROSEBUD); a TPC sensitive to the
direction of the nuclear recoil (DRIFT); devices which use
superheated droplets (SIMPLE and PICASSO), or those which use
colloids of superconducting superheated grains (ORPHEUS).

There exist also projects featuring a large amount of target
nuclei in segmented detectors, both with ionization Ge detectors
(GENIUS, GEDEON) and cryogenic thermal devices (CUORE). Table
\ref{tabla1} gives a rough account of the "history" of WIMP
searches. Table \ref{tabla2} gives an overview of the experiments
on direct detection of WIMPs currently in operation or in
preparation. General reviews on WIMPs can be found in Ref
\cite{Gri} whereas neutralino dark matter has been extensively
described in Ref \cite{Bottino} (see also the A. Bottino talk in
these Proceedings). WIMP direct detection are reviewed, for
instance in Ref \cite{Mor3}.

\begin{table*}[ht]
\caption{WIMP direct searches history} \label{tabla1}
\footnotesize
\begin{center}
\begin{tabular}{clcl} \hline
\multicolumn{2}{c}{Ge IONIZATION
      DETECTORS}&\multicolumn{2}{c}{SCINTILLATORS}\\
\hline
\\
 1986&    USC-PNNL (Homestake)&1990&    ZARAGOZA NaI (Canfranc)\\
 &UCSB-LBL (Oroville)& &ROMA lqXe (Gran Sasso) \\
 &ZAR-USC-PNL (Canfranc)& & ROMA/SACLAY NaI (LSM/LNGS)\\
&CALTECH-PSI-N (Gothard)& & UKDMC NaI (Boulby)\\
1990&H/M (Gran Sasso)& &DAMA NaI, CaF$_{2}$ (LNGS) \\
&IGEX (Canfranc)& & SACLAY NaI (Frejus)\\
&COSME (Canfranc)& & ELEGANTS NaI, CaF$_{2}$ (Oto)\\
&TAN-USC-PNL-ZAR (Sierra G)&2000 &ZEPLIN Xe (Boulby) \\
&IGEX (Baksan)& &NAIAD NaI (Boulby) \\
&HDMS (Gran Sasso)&2001 & ANAIS NaI (Canfranc)\\
2002&GENIUS-TF (Gran Sasso)& 2002& LIBRA NaI (LNGS)\\
\\
\hline
\multicolumn{2}{c}{THERMAL DETECTORS (PHONONS)}&\multicolumn{2}{c}{CRYO-DET (PHON+IONIZ)}\\
\hline
\\
 1988&MIBETA TeO$_{2}$ (LNGS) &1988 & CDMS-I Si/Ge (SUF)\\
 &EDELWEISS-0 Al$_{2}$O$_{3}$ (Frejus) & 90's&EDELWEISS I Ge (Frejus) \\
 90's &CRESS-I Al$_{2}$O$_{3}$ (LNGS) &2001 & EDELWEISS II (Frejus)\\
 & ROSEBUD Al$_{2}$O$_{3}$/Ge (Canfranc)& 2002&CDMS-II Ge/Si (Soudan) \\
 2002& CUORICINO TeO$_{2}$ (LNGS)& & \\
 2005&CUORE TeO$_{2}$ (LNGS) & & \\
 \\
\hline
\multicolumn{2}{c}{CRYO-DET (PHONONS+LIGHT)}&\multicolumn{2}{c}{SUPERC. SUPERHEATED DET.}\\
\hline
\\
 2000&CRESS-II CaWO$_{4}$ (LNGS) & & R+D of SSD since 80's\\
 2001&ROSEBUD CaWO$_{4}$ and BGO (Canfranc) & &Paris, Munich, Garching, Bern, \\
 & & &Zaragoza, Oxford, Lisbon \\
 & & 2001& ORPHEUS Sn (Bern UF)\\
 \\
\hline
\multicolumn{2}{c}{TPC}&\multicolumn{2}{c}{SUPERH. DROPLET DET. SDD}\\
\hline
\\
 2002& DRIFT Xe (Boulby)&1997 & SIMPLE Freon (Rustrel)\\
 2005& &1997 & PICASSO Freon (SNO)\\
\\
 \hline
\end{tabular}
\end{center}
\end{table*}
 \normalsize

\begin{table*}[ht]
\caption{WIMP Direct Detection in underground facilities
experiments currently running (or in preparation) } \label{tabla2}
\footnotesize
\begin{center}
\begin{tabular}{lll}
\hline LABORATORY &  EXPERIMENT & TECHNIQUE \\ \hline
BAKSAN (Russia) & IGEX &  3$\times$1 kg Ge-ionization \\
BERN(Switzerland) &  ORPHEUS &(SSD) Superconducting Superheated Detector, 0.45 kg Tin \\
BOULBY & NaI & NaI scintillators of few kg (recently completed) \\
(UK)& NAIAD &  NaI unencapsulated scintillators (50 kg)\\
& ZEPLIN & Liquid-Gas Xe scintillation/ionization  I: 4 kg single phase\\
& & II: 30 kg Two phases \\
&    DRIFT &  Low pressure Xe TPC (in preparation) 1m$^{3}\rightarrow$ 10 m$^{3}$ \\
CANFRANC  &  COSME  & 234 g Ge ionization  \\
(Spain) & IGEX  &  2.1 kg Ge ionization \\
&    ANAIS &  10$\times$10.7 kg NaI scintillators \\
&    ROSEBUD & 50g Al$_{2}$O$_{3}$ and 67g Ge thermal detectors \\
& &        CaWO$_{4}$ 54g and BGO 46g scintillating bolometers \\
FREJUS/MODANE &  SACLAY-NaI &   9.7 kg NaI scintillator (recently completed) \\
(France) &   EDELWEISS I &70 g Ge thermal+ionization detector \\
 &   EDELWEISS II  &  4$\times$320 g Ge thermal+ionization detectors \\
GRAN SASSO  &H/M &2.7 kg Ge ionization \\
 (Italy) &HDMS   & 200g Ge ionization in Ge well \\
&    GENIUS-TF  & 40$\times$2.5 kg unencapsulated Ge (in preparation) \\
&    DAMA   & NaI scintillators (87.3 kg) \\
&    LIBRA  & NaI scintillators 250 kg (in preparation) \\
& Liquid-Xe &  Liquid Xe scintillator (6 kg) \\
&    CaF$_{2}$ &   Scintillator \\
&    CRESST I &   (4$\times$260g) Al$_{2}$O$_{3}$ thermal detectors \\
&    CRESST II  & Set of 300g CaWO$_{4}$ scintillating bolometers (up to 10 kg) \\
&    MIBETA  & 20$\times$340g TeO$_{2}$ thermal detector \\
&    CUORICINO &  56$\times$760g TeO$_{2}$ thermal detector (being mounted) \\
&    CUORE   &1000$\times$760g TeO$_{2}$ (in preparation) \\
RUSTREL (France)&   SIMPLE & (SDD)Superheated Droplets Detectors (Freon) \\
STANFORD UF/ & CDMS - I &   100g Si; 6$\times$165g Ge thermal+ionization detectors \\
SOUDAN(USA)&   CDMS - II &  3$\times$250g Ge and 3$\times$100g Si Thermal+Ionization \\
SNO (Canada) & PICASSO &(SDD)Superheated Droplets Detectors (1.34g of Freon) \\
OTO & ELEGANTS-V &   Large set of massive NaI scintillators \\
 (Japan) &ELEGANTS-VI &  CaF$_{2}$ scintillators \\
 \hline
\end{tabular}
\end{center}
\end{table*}
\normalsize

\section{Strategies for WIMP direct detection}

Noticed first that the smallness of the predicted rate (which goes
from $R\sim1-10$ c/kg.day down to $10^{-4}-10^{-5}$ c/kg.day)
implies that the sensitivity of the experiment must be driven to
the best possible achievable value in this range. The
$\sigma_{\chi N}$ calculations are made within the Minimal
Supersymmetric extension of the Standard Model, MSSM, as basic
frame, implemented in various schemes (see Ref.\cite{Bottino}).
Besides the peculiarities of the SUSY model there is a wide choice
of parameters entering in the calculation of the rates: the halo
model, the values of the parameters in the WIMP velocity
distribution, the three levels of the WIMP-nucleus interaction
(quark-nucleon-nucleus) and the constraint of getting the proper
relic abundance of the candidates, just to mention a few.

From all these ingredients it follows that the theoretical
prediction of the rates show considerable spreading and are
presented as "scatter plots" extending along the various orders of
magnitude quoted above. Some of the most favorable predictions are
already testable by the leading experiments which have, in fact,
penetrate into the scatter plot of predictions. The bottom of the
plot is still far away of the detector sensitivity. It should be
noted, however, that most of the experimental searches for
SUSY-WIMPs concentrate in the dominant, coherent interaction,
which provides the largest signals but our scarce knowledge of the
nature of the WIMP interaction makes other options non-negligible.

The rarity and smallness of the signals dictate the obvious
strategy: to use ultra-low background detectors of the lowest
possible energy threshold plus one (or various) unambiguous
background rejection mechanisms, all these prescriptions carried
out in a radioactivity-free environment (including shielding,
experimental devices, ...). Example of low background recently
achieved is the case of IGEX, with Ge ionization detectors; the
cases of the CDMS and EDELWEISS which use Ge thermal detectors
which also measure ionization to discriminate, and that of ZEPLIN
which uses background discrimination in liquid Xenon.

Examples of low energy threshold and high efficiency detectors are
the bolometer experiments (MIBETA, CRESST, ROSEBUD, CUORICINO,
CDMS and EDELWEISS) which seen efficiently the energy delivered by
the WIMP (quenching factor essentially unity) and which have
achieved very low energy thresholds ($E_{THR}\sim$ hundreds of
eV).

To search for such rare events large masses of targets are also
recommended, to increase the probability of detection and the
statistics. The DAMA, UKDMC and Zaragoza scintillations
experiments use from 50 to 100 kg of NaI; the CUORICINO bolometer
experiment is installing 39.5 kg of TeO$_{2}$ crystals and the
ZEPLIN detector which uses (according to the various versions) 20
to 40 kg of Xe. It is remarkable that small size, first generation
detectors have reached exclusions $\sigma_{\chi
p}>10^{-5}-10^{-6}~$pb$~(10^{-41}-10^{-42}~$cm$^{2})$ in the range
of masses relevant for SUSY-WIMPs.

The basic idea behind the background rejection techniques is to
discriminate first electron recoils (tracers of the background)
from nuclear recoils (originated by WIMPs and neutrons). Methods
used to discriminate backgrounds from nuclear recoils are either
simply statistical, like a pulse shape analysis, PSD (based on the
different timing behavior of both types of pulses), or on an event
by event basis by measuring simultaneously two different
mechanisms of energy deposition having different responses for
background and signals, like the ionization (or scintillation) and
the heat produced by the WIMP-induced nuclear recoil, and
capitalizing the fact that for a given deposited energy (measured
as phonons) the recoiling nucleus ionizes less than the electrons.
Examples of PSD are the sodium iodide experiments of UKDMC,
Saclay, DAMA and ANAIS. Notable discrimination is also obtained in
liquid scintillators like the ZEPLIN detectors. Event by event
discrimination has been successfully applied, for instance, in
CDMS and EDELWEISS by measuring ionization and heat and in CRESST
and ROSEBUD by measuring light and heat.

Another discriminating technique is that used in the two-phase
liquid-gas Xenon detector with ionization plus scintillation, of
the ZEPLIN series of detectors. An electric field prevents
recombination, the charge being drifted to create a second pulse
in addition to the primary pulse. The amplitudes of both pulses
are different for nuclear recoils and electrons and that allows
their discrimination.

One could use instead threshold detectors--like neutron
dosimeters-- which are blind to most of the low Linear Energy
Transfer (LET) radiation (e, $\mu$, $\gamma$) and so able to
discriminate gamma background from neutrons (and so WIMPs).
Detectors which use superheated droplets which vaporize into
bubbles by the WIMP (or other high LET particles) energy
deposition are those of the SIMPLE and PICASSO experiments. An
ultimate discrimination will be the identification of the
different kind of particles by the tracking they left in, say, a
TPC, plus the identification of the WIMP through the directional
sensitivity of the device (DRIFT). Intense R+D programs are
underway to use devices with such kind of sensitivity. Tables
\ref{tabla3} to \ref{tabla7} gives, synoptically the main
non-cryogenic experiments currently running, summarizing some of
their features. Section 5 to 9 will describe and comment the
results of some of these experiments.

\section{Excluding WIMPs of the DM budget}

The direct experiments measure the differential event rate (energy
spectrum) in the customary differential rate unit (dru)
$\left[\frac{dR}{dE_{\mathrm{VIS}}}\right]^{\mathrm{exp}}$(c/keVkgday).
The registered counts contain the signal and the background. Then,
by applying discrimination techniques one disentangle at least
partially the nuclear recoils from the background events. The
resulting residual rate is then compared (in terms of m, and
$\sigma_{\chi N}$) with the theoretical nuclear recoil rate due to
WIMPs interaction first derived by Goodman and Witten
\cite{Goodman}.

The differential rate is expressed as
$$\frac{dR}{dE_{R}}=N\frac{\rho}{m}\int_{V_{min}}^{V_{esc}}\frac{d\sigma(v,E_{R})}{dE_{R}}\bar{v}f(\bar{v})d^{3}v$$
where $f(\bar{v})$ is the velocity distribution of WIMPs in the
laboratory frame and $d\sigma/dE_{R}$ the differential
cross-section WIMP-nucleus.  By assuming dominance of coherent
cross-section one has
$$\frac{dR}{dE_{R}}=N\frac{\rho}{m}\frac{(m+M)^{2}}{2m^{2}M} \sigma_{\chi N}^{SI}F^{2}(q)I(E)$$
where F is the nuclear form factor and I the integral
$I\equiv\int_{V_{min}}^{V_{esc}}\frac{f(\bar{v})}{v}d^{3}v$.
$\sigma_{\chi N}^{SI}$ is the cross-section WIMP-nucleus (of the
detector) and is usually parameterized in terms of the
WIMP-nucleon cross-section. For the standard halo model
(isothermal sphere), currently used as a first approximation to
the galactic halo, the velocity distribution function reads,
$$f(\bar{v})d^{3}v=\left({\frac{\sqrt{3/2\pi}}{v_{rms}}}\right)^{3}exp\left\{{-\frac{3(\bar{v}+{\bar{v}}_{E})^{2}}{2v_{rms}^{2}}}\right\}d^{3}v$$
with $v_{rms} \sim\sqrt(\frac{3}{2}).v_{Sun}$

$$\left[\frac{dR}{dE_{\mathrm{VIS}}}\right]^{\mathrm{Th}}=7.76\times10^{14}\frac{N}{Q}\frac{\rho}{v_{E}}\frac{(m+M)^{2}}{4m^{3}M}F^{2}\sigma^{SI}_{\chi
N}\tau$$ (with $\sigma$ in $cm^{2}$, m and M in GeVs, $v$ in
$Kms^{-1}$, $\rho$ in $GeVcm^{-3}$ and N in $kg^{-1}$) where
$\rho$ is the local density of WIMP, N the number of target
nuclei, $F^{2}$ is the nuclear form factor, and
$\tau(v_{\mathrm{esc}}\rightarrow\infty)=erf(x+y)-erf(x-y)$, with
$$x,y=\sqrt{\frac{3}{2}}(v_{\mathrm{min}},v_{E})\frac{1}{v_{\mathrm{rms}}}~~,~~~~v_{\mathrm{rms}}\sim\sqrt{\frac{3}{2}}v_{\mathrm{sun}}$$
$v$ the WIMP velocity (Earth/Lab frame) and $v_{E}$ the velocity
of Earth/Solar system with respect to the halo.

$v_{min}(E_{R})=\frac{m+M}{m}(E_{R}/2m)^{\frac{1}{2}}$ is the
minimal velocity to produce a recoil $E_{R}$. The spin-independent
nuclear cross-section is usually normalized in terms of that on
nucleons
$$\sigma^{SI}_{\chi N}=A^{2}\frac{\mu^{2}_{\chi N}}{\mu^{2}_{\chi n}}\sigma^{\mathrm{scalar}}_{\mathrm{nucleon(p.n)}}$$

Those values of ($\sigma$, m) predicting a recoil spectrum above
the observed rate
$$\left[\frac{dR}{dE_{\mathrm{VIS}}}\right]^{\mathrm{Th}}\geq\left[\frac{dR}{dE_{\mathrm{VIS}}}\right]^{\mathrm{exp}}_{\mathrm{UppBound}}$$ are
excluded. The region above the contour $\sigma(m)$ is depicted as
an exclusion plot of those WIMPs of mass $m$ with interaction
cross-section above $\sigma$. Obviously, the smaller the
background the better the exclusion.

\section{The identification of WIMP dark matter}

After reducing maximally the background and extremating the
discrimination (99.99 \%) of the detector, one should look for
asymmetries characteristic of WIMP signals. Typical smoking guns
of WIMPs could be the annual modulation of the rate
\cite{Drukier}, the forward/backward asymmetry of the nuclear
recoil \cite{Spergel} or the nuclear target dependence of the
rates \cite{Smith}.

The two kinematical asymmetries characteristic of WIMPs signals
are originated by the Earth motion through the galactic halo. The
Earth orbital motion around the Sun has a summer/winter variation,
which produces a small annual modulation of the WIMP interaction
rates, of the order
$O\left(\frac{v_{\mathrm{rot,E}}}{v_{h}}\right)\sim\frac{15}{270}\sim5\%$
\cite{Drukier}. The observation of a tiny modulation of a very
small signal requires large target mass and exposure, superb
stability and extreme control of systematics and of other
stational effects.

The orbital velocity of Earth around the sun is of 30 $Kms^{-1}$
in an orbit inclined $\alpha=60^{\circ}$ with respect to the
galactic disk
$$v_{E,r}=30\cos\alpha\cos\omega(t-t_{0})\rightarrow15\cos\omega(t-t_{0})Kms^{-1}$$
$$\omega=\frac{2\pi}{T}~~~~~T=1 \mathrm{year}~~~~t_{0}:
\mathrm{June}~2^{nd}$$ So, the velocity of the Earth (and of our
earthborn detector) relative to the galactic halo is
$$v_{E}=v_{\mathrm{sun}}+15\cos\omega(t-t_{0})Kms^{-1}$$
Consequently, in summer there is a component of the Earth' motion
around the sun parallel to the sun motion through the galaxy which
adds 15 $Kms^{-1}$. On the contrary, in winter the same occurs but
the motion is antiparallel and so one has to subtract 15
$Kms^{-1}$. The result is that the detector moves slightly faster
in June than in December (5\% effect), and consequently a
modulation of the WIMP interaction rates follows, given at first
order by $$S(t)=S_{0}+S_{m}\cos\omega(t-t_{0})[+B]$$ where $S_{0}$
is the average signal amplitude, $S_{m}$ the modulated amplitude
and $B$ the constant background.

The annual modulation signature has been already explored.
Pioneering searches for WIMP annual modulation signals were
carried out in Canfranc (NaI-32)\cite{NaI-32}, Kamioka
(ELEGANTS)\cite{AntiguaKamNaI} and Gran Sasso
(DAMA-Xe)\cite{DAMA-Xe}. In July 1977 the DAMA experiment at Gran
Sasso, using a set of NaI scintillators, reported an annual
modulation effect which after four yearly periods\cite{Ber99} has
a $3\sigma$ level significance. Such effect is compatible with the
seasonal modulation rate which could be generated by a WIMP and
so, it has been attributed by the DAMA collaboration to a WIMP of
about 60 GeV of mass and of scalar cross-section on protons of
$\sigma_{\rm p} = 7 \times 10^{-6}$ picobarns.

A second characteristic signature of the WIMP is provided by the
directional asymmetry of the recoiling nucleus \cite{Spergel}. The
WIMPs velocity distribution in the Earth frame is peaked in the
opposite direction of the Earth/Sun motion through the halo, and
so the distribution of nuclear recoils direction shows a large
asymmetry forward/backward (F/B) not easily mimicked by the
supposedly isotropic background. The order of magnitude of the
effect is large because the solar system' motion around the
galactic center $v_{\mathrm{sun}}$, and the typical WIMP velocity
in the halo, $v_{h}$, are of the same order
$O\left(\frac{v_{\mathrm{sun}}}{v_{h}}\right)\sim\frac{230}{270}\sim1$.

The angular dependence of event rate is given by (see
Ref.\cite{Spergel}) $$\frac{d^{2}R}{dE_{R}d(\cos\gamma)}=$$
$$\frac{N\rho\sigma}{\sqrt{\pi}}\frac{(m+M)^{2}}{2m^{3}Mv_{\mathrm{halo}}}~~\mathrm{exp}\left\{-\frac{(\textit{v}_{E}\cos\gamma-\textit{v}_{\mathrm{min}})^{2}}{\textit{v}^{2}_{h}}\right\}$$
where $\gamma$ is the angle of the recoiling nucleus.

There exists an increasing interest in developing devices
sensitive to the directionality of nuclear recoils from WIMPs and,
in general, to the tracking of particles. Chambers with such
purpose are being used or planned for experiments in rare event
physics. The DRIFT (Direction Recoil Identification From Tracks)
detector is a TPC of Xe (or other gases), which is sensitive to
the directionality of the nuclear recoil. The forward-backward
asymmetry of the WIMP signal will be investigated with that
device. Information and current results of the experiments
mentioned above can be found in \cite{Mor3} and References therein
and in the Proceedings of the series of TAUP
Conferences\cite{taup}. See also the Proceedings of the Neutrino
2002 Conference.

Recently, the nuclear recoil angular dependence of WIMP
interactions has been analyzed in different halo models with the
purpose of exploring how well a directional signal can be
distinguished without ambiguity from the background -with
independence of the halo model-. Quite remarkably, if the device
has angular resolution sensitivity, few events will be enough to
distinguish the signal, and not too many are needed if it has only
F/B sensitivity (see Ref.\cite{Copi}).

Another asymmetry is the nuclear target dependence of the rate
\cite{Smith}, for instance in the nuclear mass A, or in the
nuclear spin J. However, due to the differences in the intrinsic
backgrounds of the various targets, it is not easy to get reliable
conclusions. Some experiments are operating (or could it) sets of
similarly produced crystals of different nuclear targets in the
same environment like ROSEBUD
($Ge/Al_{2}O_{3}/CaWO_{4}$)\cite{UltimoRosebud} with the objective
of exploring such nuclear target dependence.

\section{Germanium Experiments}

The high radiopurity and low background achieved in Germanium
detectors, their fair low energy threshold, their reasonable
Quenching Factor (about 25\%) (nuclear recoil ionization
efficiency relative to that of electrons of the same kinetic
energy, or ionization yield) and other nuclear merits make
Germanium a good option to search for WIMPs with detectors and
techniques fully mastered. The first detectors applied to WIMP
direct searches (as early as in 1987) were, in fact, Ge
diodes\cite{twin,Jmor,Gar92,ucsb,calt,hm,Mor-SC,Ceb00}, as
by-products of $2 \beta$-decay dedicated experiments. Table
\ref{tabla3} shows the Germanium ionization detector experiments
currently in operation. We will review however only some examples.
Pioneer germanium experiments looking for WIMPs are describes in
Refs.

\begin{table*}[ht]
\caption{Ge Ionization Experiments} \label{tabla3} \footnotesize
\begin{center}
\begin{tabular}{lcclll} \hline
Experiment&M (kg)&$E_{Thr}(keV)$&$\Gamma (keV)$&Low Energy&Observations\\
and site& & Q$\sim0.25$ & &B(c/keVkgday)& \\
\hline
 COSME-II &0.234&2.5&0.4 (at 10 keV)&0.6(2-15 keV)&Good exclusions for\\
LSC-Canfranc&(311d)& & &0.3(15-30 keV)&low mass WIMPs\\
\\
 IGEX &2.1 &4&2 (at 10 keV)&0.21 (4-10 keV)&1 detector from IGEX \\
LSC-Canfranc&(139d)& & &0.10 (10-20 keV)&enriched $^{76}$Ge set (2$\beta$)\\
&&&&0.04 (20-40 keV)\\
\\
H/M &2.76 &9&2.4 (at 727 keV) &0.16 (9-15 keV) &1 detector from H/M \\
LNGS&(250d)& &extr. 2 at 0 keV&0.042 (15-40 keV)&enriched $^{76}$Ge set(2$\beta$)\\
\\
HDMS &inn. 0.2 &2.5 &1.2 (at 300 keV)&0.2 (11-40 keV) &Small detector inside a \\
LNGS& & &extr. 1 at 0 keV& 0.07 (40-100 keV)&well-type Ge \\
 &out. 2.1 &7.5& 3.2 (at 300 keV) & &outer crystal\\
 &(49d)& &extr. 3 at 0 keV\\
\\
GENIUS-TF  &40 &0.5 nom. & &goal&14 Ge crystals embedded \\
LNGS&(2.7x14)& 12 eff.& &$10^{-2}$ & in liquid nitrogen\\
In preparation & &(cosmog.)& & (12-100 keV)&housed in zone-refined\\
 & & & & &Ge bricks\\
\\
GENIUS &100-&$>12$ & &goal &Large set of naked\\
Project& 10000&(cosmog.)& & $10^{-4}-10^{-5}$&p-type Ge
detectors\\
 & & & &(12-100 keV) & in liquid Nitrogen\\
 \\
GEDEON &56 &1-2 nom. &1 (at 10 keV)&goal&28 Ge diodes in one\\
Project & 4x7(2kg)&12 eff.& &$10^{-2}-10^{-3}$&single cryostat\\
LSC-Canfranc& &(cosmog.)& &(2-50 keV)&archaeological lead\\
 & & & & &and pure graphite\\
 \hline
\end{tabular}
\end{center}
\end{table*}
\normalsize

\begin{figure}[h]
\centerline{\includegraphics[width=7cm]{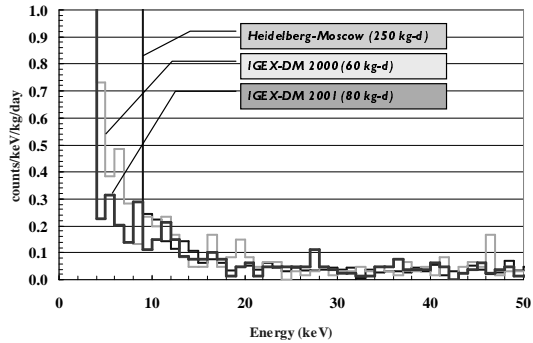}} \caption{Background
spectra of the Ge experiments, IGEX and H/M.} \label{fig1}
\end{figure}

The International Germanium Experiment (IGEX) which was optimized
to search for the double beta decay of germanium \cite{Aal,Gon99}
is using one enriched detector of $^{76}$Ge of $\sim$2.1 Kg to
look for WIMPs in the Canfranc Underground
Laboratory\cite{Mor00,Mor01}. It has an energy threshold of 4 keV
and an energy resolution of 0.8 keV at the 75~keV Pb x-ray line.
The detector is fitted in a cubic block in lead being surrounded
by not less than 40-45 cm of lead of which the innermost 25 cm are
archaeological. A muon veto and a neutron shielding of 40 cm of
polyethylene and borated water completed the set-up
\cite{Mor00,Mor01}. The spectrum of IGEX-2001 \cite{Mor01}
together with that of a previous run \cite{Mor00} (IGEX-2000) are
shown in Fig. \ref{fig1} in comparison with that of the
Heidelberg-Moscow experiment\cite{Bau99}. The H/M experiment is
another enriched-Ge experiment (enriched $^{76}$Ge crystal of 2.7
kg and energy threshold of 9 KeV), already completed, which has
been running at Gran Sasso.

The exclusion plots resulting from the IGEX data are derived from
the recorded spectrum Fig \ref{fig1} in one-keV bins from 4 keV to
50 keV. The method followed in deriving the plot has been the same
for all the detectors and experiments from which exclusion plots
are despicted in this review. As recommended by the Particle Data
Group, the predicted signal in an energy bin is required to be
less than or equal to the (90\% C.L.) upper limit of the (Poisson)
recorded counts. The derivation of the interaction rate signal
supposes that the WIMPs form an isotropic, isothermal,
non-rotating halo of density $\rho = 0.3$~GeV/cm$^{3}$, have a
Maxwellian velocity distribution with $\rm v_{\rm rms}=270$~km/s
(with an upper cut corresponding to an escape velocity of
650~km/s), and have a relative Earth-halo velocity of $\rm v_{\rm
r}=230$~km/s. The cross sections are normalized to the nucleon,
assuming a dominant scalar interaction. The Helm parametrization
\cite{Eng91} is used for the scalar nucleon form factor, and the
recoil energy dependent ionization yield used is the same that in
Ref \cite{Bau99} $\rm E_{vis} = 0.14 (E_{REC}) ^{1.19}$.

\begin{figure}[h]
\centerline{\includegraphics[width=7cm]{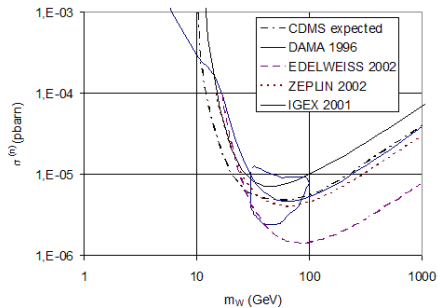}}
\caption{Exclusion plots obtained by the most advanced experiments
taking as reference their crossing of the region define by the
DAMA candidate.} \label{fig2}
\end{figure}

The best exclusion plot derived for these Ge experiments, that of
IGEX-2001, is depicted in Fig. \ref{fig2} and labelled on the
right border of the figure. IGEX-2001 improves the exclusion of
the other Ge-ionization experiments for a mass range from 20~GeV
up to 200~GeV, which encompass the DAMA mass region \cite{Ber99}.
In particular, IGEX excludes WIMP-nucleon cross-sections above 7
$\times 10^{-6}$ pb for masses of 40-60 GeV and enters the DAMA
region excluding the upper left part of this region. That is the
first time that a direct search experiment with a Ge-diode without
background discrimination, but with very low (raw) background,
enters such region. A further 50 \% background reduction between 4
keV and 10 keV would allow IGEX to explore practically all the
DAMA region in 1 kg.y of exposure.

\begin{figure}[h]
\centerline{\includegraphics[height=5.4cm]{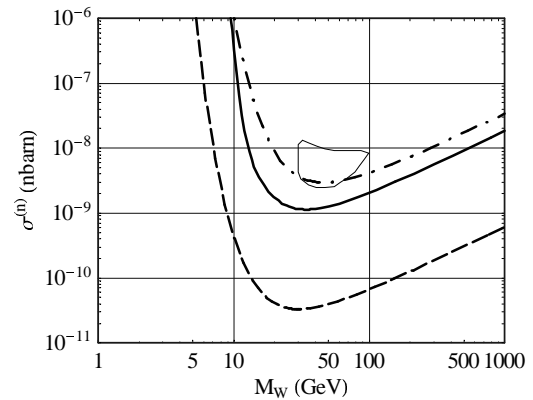}}
\caption{IGEX-DM projections are shown for
 a flat background rate of 0.1 c/keVkgday (dot-dashed line) and 0.04 c/keVkgday (solid line) down
 to the threshold at 4 keV, for 1 kg.year of exposure.
 The exclusion contour expected for GEDEON is also
 shown (dashed line) as explained in the text.}
\label{fig3}
\end{figure}

In Figure \ref{fig3} we plot the exclusions that would be obtained
by IGEX with a flat background of 0.1 c/kgkeVday (dot-dashed line)
and of 0.04 c/kgkeVday (solid line) down to the current 4 keV
threshold, for an exposure of 1 kg.year \cite{Igor-Tesis}.

In Figure \ref{fig2}, also shown for comparison are the contour
lines of the experiments which have crossed partially of totally
the DAMA region like CDMS \cite{Abusaidi:2000}, EDELWEISS
\cite{Benoit:2001} and ZEPLIN \cite{Liubarski}. The DAMA region
(closed line) corresponding to the annual modulation effect
reported by that experiment \cite{Ber99} and the exclusion plot
obtained by DAMA NaI-0 \cite{dama} using statistical pulse shape
discrimination are also shown.

There exist new experimental projects to look for WIMPs with Ge
detectors: GEDEON (GErmanium DEtectors in ONe cryostat), is
planned to use 56 kg og Ge of natural isotopic abundance
\cite{Mor-SC,Igor-Tesis,Mor992}. It will use the technology
developed for the IGEX experiment and it would consist of a set of
$\sim$2 kg Germanium crystals, of a total mass of about 56 kg,
placed together in a compact structure inside one only cryostat.
This approach could benefit from anticoincidences between crystals
and a lower components/detector mass ratio to further reduce the
background with respect to IGEX.

The GEDEON single cell is a cylindrical cryostat in electroformed
copper hosting 28 germanium crystals which share the same common
copper cryostat (0.5 mm thick). The Ge crystals, are arranged in
four plates of seven detector each suspended from copper rods. The
cell is embedded into a precision-machined hole made in a Roman
lead block providing a shield of 20 cm, and surrounded by another
lead shielding 20 cm thick. A cosmic veto and a large neutron
shield complete the shielding.

\begin{figure}[h]
\centerline{\includegraphics[height=5.4cm]{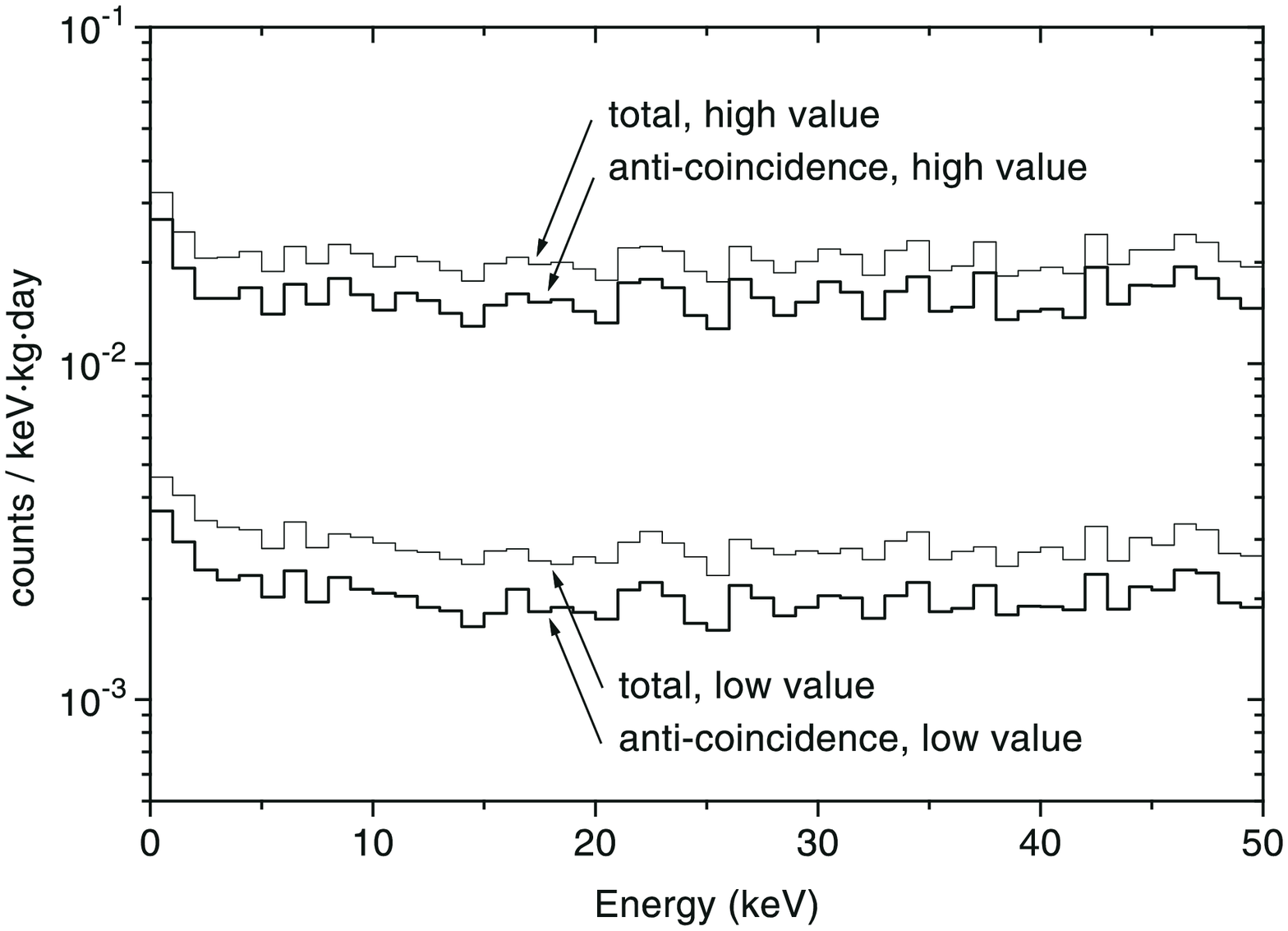}}
\caption{Montecarlo estimated background of the GEDEON detector
 project (energy interval 0---50 keV).}
 \label{fig4}
\end{figure}

The preliminary MC estimated intrinsic (assumed dominant)
background in the $1 \sim 50$ keV region ranges from $2 \times
10^{-2}$ to $2 \times 10^{-3}$ c/keVkgday, according to the level
of radioimpurities included as input \cite{SC-privada}. The
radiopurity assays have been carried out in the Canfranc
Laboratory for the lead and copper components of the shielding.
The background final goal of GEDEON, below 100 keV, would be in
the region of $10^{-3}$ c/keV.kg.day and this value has been used
to calculate anticipated $\sigma$(m) exclusion plots in the most
favourable case. The expected threshold assumed has been $\rm
E_{\rm thr}=2$ keV and the energy resolution in the low energy
region has been taken $\Gamma \sim 1$ keV. The MC estimated
intrinsic background of the GEDEON unit cell (28 crystals) is
given in Fig. \ref{fig4}. A detailed study is in progress to
assess the physics potential of this device. The exclusion plot
which could be expected with such proviso in a first step (28 kg.y
of exposure) is shown in Figure \ref{fig3}. Moreover, following
the annual modulation sensitivity plots presented in \cite{SC-01},
GEDEON would be massive enough to search for the WIMP annual
modulation effect \cite{Drukier,Freese} and explore positively an
important part of the WIMP parameter space.

Two more Ge experiments running or in preparation, both in Gran
Sasso, are that of the Heidelberg Dark Matter Search (HDMS) and
the GENIUS-Test Facility. The small detector of HDMS has achieved
a background still higher than that of H/M and so the results will
not be include here. See Ref.\cite{bau00}. Most of the attention
of this Collaboration goes now to the preparation of a small
version with natural abundance germanium detectors (GENIUS-Test
Facility) of the GENIUS project \cite{genius}. GENIUS is a
multipurpose detector, consisting of enriched Ge detectors of
about 2$\frac{1}{2}$ kg each (up to a total of 0.1 to 10 tons)
which uses the novel idea of immerse the crystals directly into a
large tank of liquid nitrogen. GENIUS-TF, now in preparation, is
intended to test the GENIUS project and at the same time to search
for WIMP.


\section{WIMP searches with NaI scintillators}

The sodium iodide detectors are very attractive devices to look
for WIMP. Both nuclei have non-zero spin
($^{23}Na~J=\frac{3}{2},~^{127}I~J=\frac{5}{2}$) and then
sensitive also to spin dependent interaction. Iodine is a heavy
nucleus favourable for spin-independent interactions. The
quenching factor is small ($Q<10\%$) for I, and medium for Na
($Q\sim30-40\%$). Backgrounds lesser than or of the order of $\sim
1$ c/keVkgday in the few keV region have been achieved. There
exists four NaI experiments running: DAMA, UKDMC, (in various
detectors and projects), ELEGANTS and ANAIS, and other in
preparation. Table \ref{tabla4} shows the main features of the
current NaI experiments.

\begin{table*}[ht]
\caption{NaI Scintillation Experiments} \label{tabla4}
\footnotesize
\begin{center}
\begin{tabular}{lccll} \hline
Experiment&M (kg)&$E_{Thr}(keV)$&B(c/keVkgday)&Observations\\
and site & & &aver. after PSD& \\
\hline
 UKDMC &2-10&4&2-4 (DM4G, 5 kg)&Anomalous fast event found \\
 Boulby&1.7 p.e./keV& & &not yet fully understood\\
\\
 DAMA&9 x 9.70 &2&$1\sim1.5$ (at 2-3 keV) &Annual modulation effect reported
along four \\
 LNGS& 7 p.e./keV& &$1.5\sim2$ (at 3-6 keV)& annual cycles (4$\sigma$). (5th and 6th cycles soon)\\
 &(best crystal) & & 0.5 (at 2.5 keV) &Phys. Lett. B450 (99)448; ibid B480 (2000) 23\\
 \\
ELEGANT&730&4-5&8-10 (at threshold)&Old set-up upgraded\\
Oto Cosmo&20 Modules& & & Large BKG from $^{210}$Pb (10 mBq/kg)\\
 & & & &Analysis only of 9 modules (328.5 kg)\\
 \\
 ANAIS&prototype & 2 & 3-4 (at 2 keV)&107 kg intended for ann. mod. search.\\
 LSC-Canfranc& 10.7& & 2 (at 3-5 keV)&Old set upgraded plus new radiopure crystals\\
 (Prototype & Final set & &1 (at 5-8 keV) &Preliminary 1200 kg day\\
  running)&10x10.7 & & &Also R+D in NaI unencapsulated\\
\\
 NAIAD& units of& 2 & 2 (2-20 keV) &Set of NaI unencapsulated\\
 Boulby& 5-10& & expected&prototype: 5cm$\phi$ x5cm Crystal in\\
 (In prepa- & for a set of& & &polypropylene barrel 6-12 p.e./keV\\
  ration)& 40-50& & &No fast events seen\\
\\
 LIBRA & 250 & & &R+D on detector radiopurity crystals\\
 (DAMA)& & & &from ultrapure powders\\
 LNGS & & & &(In preparation)\\

 \hline
\end{tabular}
\end{center}
\end{table*}
\normalsize

\begin{figure}[h]
\centerline{\includegraphics[height=5.4cm]{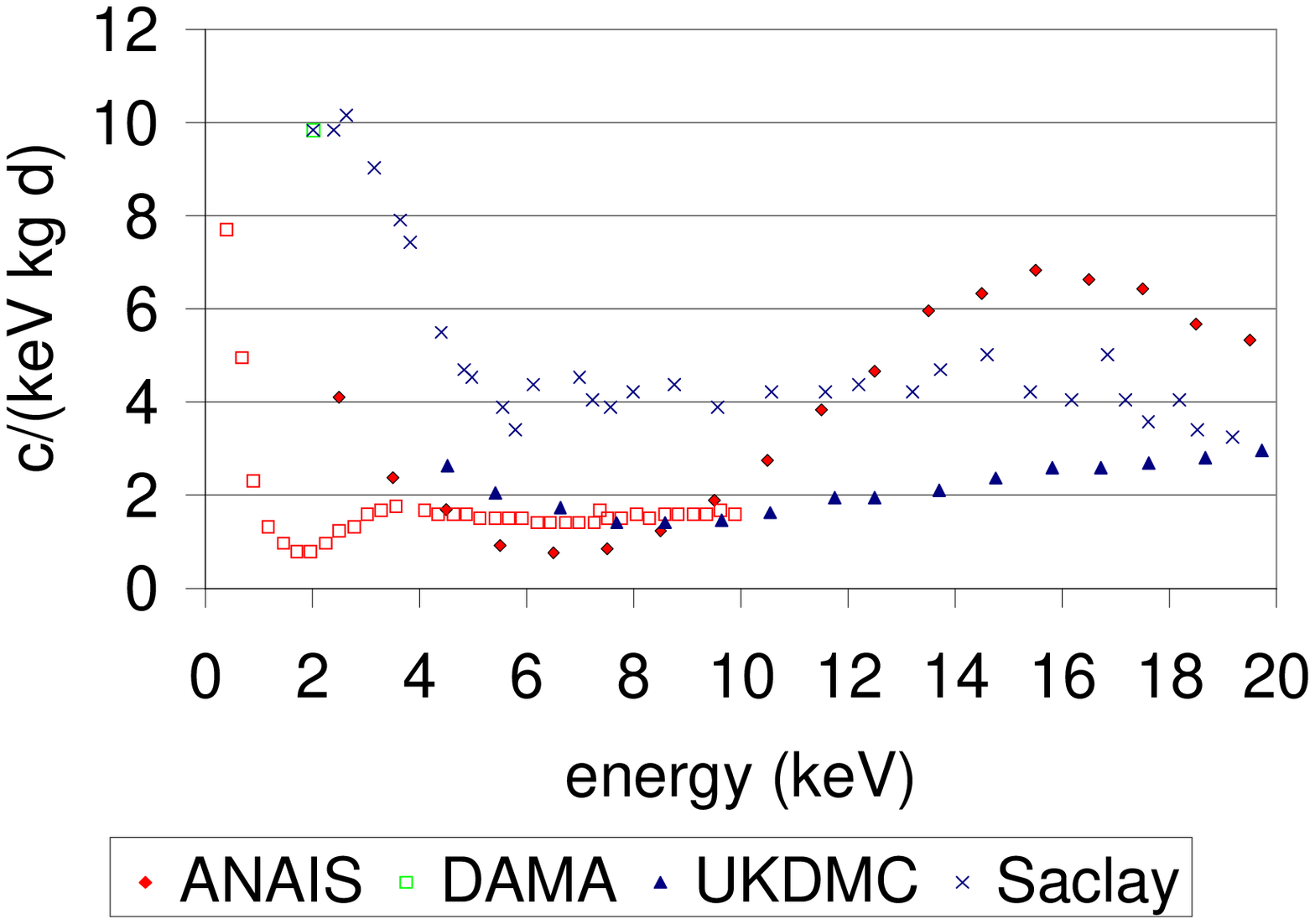}}
\caption{} \label{fig5}
\end{figure}

The NaI scintillators can be endowed with Pulse Shape
Discrimination (PSD) to distinguish statistically gamma background
from WIMPs (or neutron) signals, because of the different timing
behavior of their pulses. From such statistical analysis it
results that only a few percent (depending on the energy) of the
measured background can be due to nuclear recoils. The background
spectra (before PSD) of four NaI experiments ANAIS \cite{cebrian},
DAMA \cite{Ber99}, UKDMC \cite{ukdm} and Saclay \cite{saclay} are
shown comparatively in Fig. \ref{fig5} (1 to 2 c/keVkgday in DAMA,
UKDMC and ANAIS and of 2 to 10 c/keVkgday in Saclay and ELEGANTS
\cite{elegants}). Table \ref{tabla7} shows the typical background
reduction obtained with PSD.

\begin{table*}[ht]
\caption{Examples of PSD background rejection in NaI experiments}
\label{tabla5} \footnotesize
\begin{center}
\begin{tabular}{lcccccccl} \hline
Experiment& Exposure & Method &\multicolumn{5}{c}{PSD Background rejection 90$\%$CL} & Obs.\\
and site & &  &\multicolumn{5}{c}{(Energy interval in keV)}\\
 & & & 4-6 & 6-8 &8-10&10-12&12-20&  \\
\hline
 UKDMC&1122 kg.day &time decay  & 85$\%$&92$\%$& 94$\%$&96$\%$& 97$\%$& Phys Lett B779(96)299\\
 Boulby&181d x 6.2kg & const.& & & & & & and B433(98)150\\
 \\
 SACLAY& 850 kg.day&$1^{st}$ moment& 87$\%$&91$\%$& 92$\%$&94$\%$& 96$\%$& only stat.\\
 LSM Frejus&83d x 9.7kg & of time dist. & 65$\%$&70$\%$& 62$\%$&85$\%$& 87$\%$&incl. syst. \\
 & &(and others) & & & & & &Astrop Phys 11(99)275 \\
 \\
 DAMA 0&4123 kg.day &$1^{st}$ moment & 88$\%$&92$\%$& 96$\%$&98$\%$& 99$\%$&Phys Lett B389(96)757 \\
 LNGS& 83d x9x 9.7kg& of time dist.& & & & & & \\
 \hline
\end{tabular}
\end{center}
\end{table*}
\normalsize

The United Kingdom Dark Matter Collaboration (UKDMC) uses
radiopure NaI crystals of various masses (2 to 10 kg) in various
shielding conditions (water, lead, copper) in Boulby \cite{ukdm}.
Typical thresholds of 4 keV have been obtained. UKDMC is also
preparing, NAIAD (NaI Advanced Detector) which will consist of
50--100 kg in a set of crystals. The exclusion plots obtained with
these detectors (for spin independent couplings) are still worse
than that of Ge and will not be included here.

ANAIS (Annual Modulation with NaI's\cite{cebrian}) will use 107 kg
of NaI(Tl) in Canfranc. A prototype of one single crystal (10.7
kg) is being developed. The components of the photomultiplier have
been selected for its radiopurity. Pulse shape analysis has been
performed. The preliminary results refer to an exposure of 2069.85
$kg day$ (Fig. \ref{fig5}) shows the background after the noise
rejection. The energy threshold is of $\sim 4$ keV and the
background level registered from threshold up to 10 keV is about
1.2 c/keVkgday.

The DAMA experiment \cite{Ber99} uses 9 radiopure NaI crystals of
9.7 kg each, viewed by two PMT. The software energy threshold is
at $E_{Thr}=2$ keV and the energy resolution at 2-5 keV is
$\Gamma\sim2-2.5$ keV. The PSD method applied to the DAMA NaI-0
\cite{dama} running lead to a background reduction of 85\% (at
4--6 keV) and 97\% (at 12--20 keV), providing remarkable exclusion
plots.

\begin{figure}[h]
\centerline{\includegraphics[width=7cm]{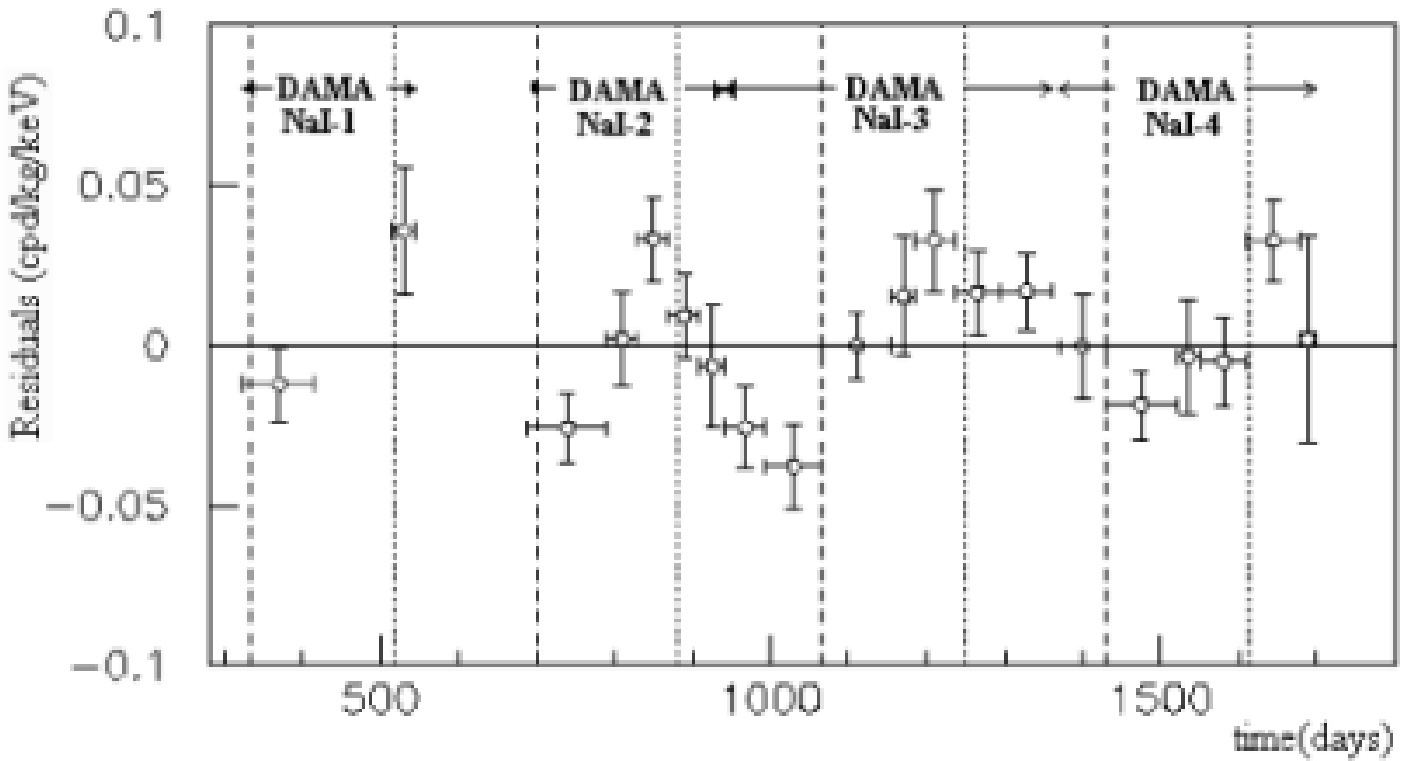}} \caption{}
\label{fig6}
\end{figure}

The main objective of DAMA however is to search for the annual
modulation of the WIMPs signal. Such modulation has been found and
attributed by the Collaboration to a WIMP signal. After 57986 kg
day of statistics the residuals of the rate vs time, looks as
shown in Fig.\ref{fig6}. It modulates according to
$A\cos[\omega(t-t_{0})]$ with period and phase consistent with 1
year and 2nd June, respectively. The probability of absence of
modulation is $\sim4\times10^{-4}$. The DAMA global results
\cite{Ber99} (NaI, 1, 2, 3, 4 running) in the case of assuming the
WIMP interpretation, lead to a WIMP of mass and cross-section
given by $M_{W}=(52^{+10}_{-8})GeV$
$\xi\sigma^{p}=(7.2^{+0.4}_{-0.9})\times10^{-6}pb$. A maximum
likelihood favours the hypothesis of presence of modulation with
the above $M_{W}$, $\xi\sigma_{p}$ values at $4\sigma$ C.L. The
($\sigma$, m) region for spin independent coupled WIMP is the
"triangle" zone depicted in Fig. \ref{fig2}. An extension of DAMA
up to 250 kg of NaI (LIBRA) is being prepared.

The DAMA results have aroused great interest and controversies. It
is imperative to confirm the DAMA results by other independent
experiments with NaI (like LIBRA and ANAIS) and with other nuclear
targets, say Ge and Te. For instance,
CUORICINO\cite{cuore,cuore2}, with 39.6 kg of TeO$_{2}$, is now
being mounted. GEDEON and GENIUS--TF of about 56 and 40 kg of
germanium respectively will critically explore that modulation
(see Ref \cite {SC-01}). The DAMA $\sigma$(m) region is being
explored, also by the standard method followed for excluding
WIMPs. Various experiments have already reached and even
trespassed the DAMA region (see in the Fig \ref{fig2} the
comprehensive exclusion plot including IGEX, CDMS, EDELWEISS and
ZEPLIN).

The OSAKA group is performing a search with the ELEGANTS V NaI
detector in the underground facility of Oto. ELEGANTS
\cite{elegants} uses huge mass of NaI scintillators (760 kg)
upgraded from a previous experiment. The background at threshold
is still high. A search for annual modulation did not show any
indication of modulation.

\section{WIMP searches with Xenon Scintillation Detectors}

The search for WIMPs with Xenon scintillators benefits of a
well-known technique. Moreover, background discrimination can be
done better than in NaI. They are also targets of heavy nuclear
mass ($A\sim130$) for enhancement of the spin-independent coherent
interaction.

They have achieved a fairly good level of radiopurity, have a good
quenching factor (Q$\sim50$\%) and a high density ($\sim$
3gr/cm$^{3}$). Summing up these properties, one conclude that
Xenon scintillators based detectors are a good option to look for
WIMPs.

One of the pioneer searches using Xenon is the DAMA liquid-Xenon
experiment. The spectra of limits on recoils in WIMP-$^{129}$ Xe
elastic scattering using PSD and exclusion plots were published in
Ref.\cite{Ber98}. Recent results of the DAMA liquid Xenon
experiment refers to limits on WIMP-$^{129}$ Xe inelastic
scattering \cite{Ber00}.

The ZEPLIN Program \cite{spooner} uses a series of Xenon-based
scintillators devices able to discriminate the background from the
nuclear recoils in liquid or liquid-gas detectors in various ways.
Either using the Scintillation Pulse Shape or measuring the
scintillation and the ionization (an electric field prevents
recombination, the charge being drifted to create a second
scintillation pulse), and capitalizing the fact that the primary
(direct) scintillation pulse and the secondary scintillation pulse
amplitudes differ for electron recoils and nuclear recoils, the
secondary scintillation being smaller for nuclear recoils. That
feature provides a powerful background rejection. The secondary
scintillation photons are produced by proportional scintillation
process in liquid-Xenon like in the ZEPLIN-I detector, (where a
discrimination factor of 98\% is achieved) or by electro
luminescence photons in gas-Xenon (like in the case of the
ZEPLIN-II detector prototype) in which the electrons (ionization)
are drifted to the gas phase where electroluminescence takes place
(the discrimination factor being $\mathrm{>99\%}$) \cite{zeplin}.
Some prototypes leave been tested and various different projects
of the ZEPLIN series are underway \cite{spooner,cline} in Boulby.
A recent running has provided a remarkable exclusion plot (see
\cite{Liubarski}) which traverse entirely the DAMA region as shown
in Fig \ref{fig2}. Table \ref{tabla6} is an sketch of the various
Xenon experiment or projects.

\begin{table*}[ht]
\caption{Xenon scintillation experiments} \label{tabla6}
\footnotesize
\begin{center}
\begin{tabular}{llccllcl} \hline
Experiment& Type of&Mass&$E_{Thr}$&Meth. of &Discr. eff./ &  BKG at Thr& Observations \\
and site & detector& (kg) & (keV)&discrim&Reject. factor & before
PSD& \\
\hline
 DAMA-Xe& Liquid Xe& 6.5& 13&PSD & $50\%$(13-15 keV)& 0.8&Classical search \\
 LNGS& & & & & $95\%$(16-20 keV)&c/keVkgd &for WIMPs with \\
 & & & & & & & Xe (since 1993)\\
 & & & & & & &Phys Lett B389\\
 & & & & & & &(96)757\\
  \\
 ZEPLIN I&Liquid Xe & 4& & PSD& & & Preliminary\\
 Boulby& & & & SCI/ION& & & \\
 \\
 ZEPLIN II & Two-Phase& 20-40 & 10&SCI/ION &$>99\%$ & goal&$1 m^{3}$ by 2001-2002\\
 Boulby &Xe & & & & & $10^{-2}$c/kg.d&1 kg prototype \\
 In prep. &Liquid-Gas & & & & & &UCLA-Torino \\
 Start 2003? & & & & & & & Low field\\
\\
 ZEPLIN III & Two-Phase&6 & &SCI/ION & & & High field 20 kV\\
 Boulby & Xe& & &improved & & & low threshold\\
 Proj. 2004 & & & & & & & \\
 \\
 ZEPLIN IV &Two-Phase &1000 & &SCI/ION & &goal& Ext. of Z II\\
 Boulby &Liquid-Gas & & & & & $10^{-4}$c/kg.d& \\
 Proj. 2006 & & & & & & & \\
 \\
 KAMIOKA & Liquid-Gas& 1& & & $99\%$& & \\
 Xe &Xe & & & & (10-100 keV)& & \\

 \hline
\end{tabular}
\end{center}
\end{table*}
\normalsize
\begin{table*}[ht]
\caption{Other techniques} \label{tabla7} \footnotesize
\begin{center}
\begin{tabular}{llccllcl} \hline
Experiment& Detector&$E_{Thr}$&Discr./ & Low Energy& Observations \\
and site & and Mass& (keV) & Reject.&Bacground &  \\
\hline
DRIFT& Xenon TPC& &$99.9\%(\gamma)$ &goal & Low pressure (10-80 Torr) \\
Boulby& $1 m^{3} to 10 m^{3}$& &$95\%(\alpha)$(wires) &$<10^{-2,-3}$c/kg.day &Xenon Negative Ions TPC\\
& & & & &to detect nuclear recoil track \\
& & & & &sensitive to directionality \\
\\
SIMPLE& Freon $C_{2}ClF_{5}$& 7& &goal & Superheated droplet detector\\
Rustrel&9.2 g R115 & & & $\sigma_{p}^{SD}\sim10^{-1,-2}$pb& Blind to typical radioactivity\\
& 1999: 0.19 kg.d& & & & (Low LET particles)\\
& 2001: 0.77 kg.d& & & & Results 1999 $\sigma_{p}^{SD}\sim5-10$pb\\
& & & & & m$\sim$(10-100 GeV) \\
\\
PICASSO&1.34 g Freon & & & & Superheated Droplet Detector\\
Sudbury&117 day data & & & &Results \\
& & & & &$\sigma_{p}^{SD}\sim10$pb; m$\sim$(10-100 GeV) \\
\\
ORPHEUS& 450 g& &97$\%$ & &Superconducting \\
Bern&Sn granules & & flip of& &superheated detectors \\
Shallow depth&$\sim30\phi \mu$m & & various grains& & \\
 \hline
\end{tabular}
\end{center}
\end{table*}
\normalsize

\section{WIMP searches with Time Projection Chambers}

DRIFT is a detector project sensitive to directionality
\cite{drift}. It uses a low pressure (10-40 Torr) TPC with Xenon
to measure the nuclear recoil track in WIMP-Nucleus interactions.
The direction and orientation of the nuclear recoil provide a
characteristic signature of the WIMP. The diffusion constrains the
track length observable but DRIFT reduces the diffusion
(transversal and longitudinal) using negative ions to drift the
ionization instead of drift electrons: gas CS$_{2}$ is added to
capture electrons and so CS$^{(-)}_{2}$ ions are drifted to the
avalanche regions (where the electrons are released) for multiwire
read-out (no magnetic field needed). The negative ion TPC has a
millimetric diffusion an a millimetric track resolution. The
proof-of-principle has been performed in mini-DRIFTs, where the
direction and orientation of nuclear recoils have been seen. The
event reconstruction, the measurement of the track length and
orientation, the determination of dE/dx and the ionization
measurement permit a powerful background discrimination (99.9\%
gamma rejection and 95\% alpha rejection) leading to a rate
sensitivity of $R<10^{-2(-3)}$c/kgday. DRIFT will permit to
recognize the forward/backward asymmetry and the nuclear recoils
angular distribution, which, as already noted, are the most clear
distinctive signatures of WIMPs. That will permit hopefully the
identification of WIMP. A DRIFT prototype of 1 m$^{3}$ is almost
completed \cite{spooner}. A project of 10 m$^{3}$ (Xe) scaling up
the TPC of 1 m$^{3}$ (Xe) is under way.

\section{WIMP searches with metastable particle detectors}

WIMP detectors, which use the metastability of the medium where
the nuclear targets are embedded, are the (novel) superheated drop
detectors (SDD) (like SIMPLE and PICASSO) and the (old)
superconducting superheated grains (SSD) (like ORPHEUS). The SDD's
consist of a dispersion of droplets ($\oslash\sim10\mu m$) of
superheated liquid (freon) in a gel matrix. The energy deposition
of a WIMP in the droplets produces a phase transition from the
superheated to normal state causing vaporization of droplets into
bubbles ($\oslash\sim 1mm$), detected acoustically. SSD are
essentially insensitive to low LET particles (e, $\gamma$, $\mu$),
and so good for detecting WIMPs and neutrons. See
Ref.\cite{simple} and \cite{picasso} for PICASSO.

The Superconducting Superheated Grains (SSG) detectors, like
ORPHEUS, is based on the change of phase from the superconducting
superheated to the normal state produced by the WIMP energy
deposition in micrograins inside a magnetic field, at very low
temperatures. The signal is detected through the disappearance of
the Meissner effect. The SSG offer good background rejection
(97\%) (a single grain is expected to flip per WIMP or nucleon
interaction, in contrast to several grains in the case of other
particles), and are sensitive to very low energy deposition (as
proved in neutron irradiation experiments). An experiment with tin
micrograins has just started at the Bern Underground Laboratory
(70 m.w.e.)\cite{orpheus,pretzl}.

\section{Conclusions and outlook}

The direct search for WIMP dark matter proceeds at full strength.
There are many experiments and projects on direct detection going
on. Only a sample of them has been chosen to illustrate the
strategies, methods and achievements. The new experiments are
focusing in the identification of WIMPs, discriminating the
nuclear recoils from the background (rather that in constraining
or excluding their parameters space) and looking for distinctive
WIMP signals.

Taking also into account the results and prospectives of the
direct detection with cryogenic/thermal devices, reviewed by L.
Mosca in these Proceedings, the present experimental situation can
be summarized as follows: the rates predicted for SUSY-WIMPs
extend from 1-10 c/kgday down to $10^{-4}-10^{-5}$ c/kgday, in
scatter plots, obtained within MSSM as basic frame implemented in
various alternative schemes. A small fraction of this window is
testable by some of the leading experiment. The rates
experimentally achieved stand around 1 c/kgday (0.1 c/kgday at
hand) (CDMS, EDELWEISS) and differential rates $\sim0.1-0.05$
c/keVkgday have been obtained by IGEX and H/M, in the relevant low
energy regions. The deepest region of the exclusion plots achieved
stands around a few $\times10^{-6}-10^{-7}$pb, for masses 50-200
GeV (EDELWEISS and ZEPLIN). The current status of the most
relevant exclusion plots (IGEX, DAMA, CDMS, EDELWEISS, ZEPLIN) is
depicted comparatively in Fig \ref{fig2}. On the other hand, there
exists an unequivocal annual modulation effect (see Fig.
\ref{fig4}) reported by DAMA (four yearly periods), which has been
shown to the compatible with a neutralino-WIMP, of m$\sim50-60$GeV
and $\sigma^{Si}_{n}\sim 7\times10^{-6}$pb. Recent experiments
exclude at greater or lesser extend (CDMS, EDELWEISS, ZEPLIN,
IGEX) the DAMA region in the case of an assumed purely scalar
interaction.

\begin{table*}[ht]
\caption{WIMP Direct Detection Prospect}
\label{tabla8}\footnotesize
\begin{center}
\begin{tabular}{ll}
\hline
 &BEING INSTALLED/OR PHASE II EXPERIMENTS~(To start 2001-2002)\\
 \hline
CDMS-II & (Ge,Si) Phonons+Ioniz 7 kg, B$\sim 10^{-2}-10^{-3}$ c/kgd, $\sigma \sim 10^{-8}$ pb \\
EDELWEISS-II & (Ge) Phonons+Ioniz 6.7 kg, B$\sim 10^{-2}-10^{-3}$ c/kgd, $\sigma \sim 10^{-8}$ pb (40-200 GeV) \\
CUORICINO & TeO$_{2}$ Phonons 42 kg, B$\sim 10^{-2}$ dru, $\sigma\sim 0^{-7}$ pb \\
CRESST-II & CaWO$_{4}$ Phonons+light, $B<10^{-2}-10^{-3}$ dru (15 keV), $\sigma \sim 10^{-7}-10^{-8}$ pb \\ & (50-150 GeV) \\
IGEX &  Ge Ioniz 2.1 kg, B$< 10^{-1}-10^{-2}$ dru, $\sigma \sim2\times 10^{-6}$ pb (40-200 GeV) \\
HDMS &  Ge Ioniz 0.2 kg, $\sigma \sim 6 \times 10^{-6}$ pb (20-80 GeV)\\
ANAIS & NaI Scintillators 107-150 kg, B(PSD)$\leq 0.1$ dru, $\sigma \sim 2 \times 10^{-6}$ pb \\
NAIAD & NaI Scintillators 10-50 kg, B(PSD)$\leq 0.1$ dru, $\sigma
\sim 10^{-6}$ pb (60-200 GeV) \\
\hline

& IN PREPARATION~(To start 2002-2003)  \\ \hline
LIBRA (DAMA) &  NaI Scintillators 250 kg \\
GENIUS-TF & Ge Ioiniz 40 kg, B$<10^{-2}$ dru, $E_{Thr}$=10 keV
$\rightarrow \sigma \sim 10^{-6}$ pb (40-200 GeV), \\
& $E_{Thr}$=2 keV $\rightarrow \sigma \sim 10^{-7} pb$ (20-80 GeV)\\
ZEPLIN-II & Xe-Two-phase 40 kg, NR discrim$>$99\%, B$<10^{-2}$
dru, $\sigma \sim 10^{-7}$ pb \\
DRIFT-I  &  Xe TPC 1 m$^{3}$, B$<10^{-2}$ dru, $\sigma \sim
10^{-6}$ pb (80-120 GeV) \\ \hline

 & IN PROJECT~($>$2003-2005) \\ \hline
CUORE & TeO$_{2}$ Phonons 760 kg, $E_{Thr}\sim$2.5 keV, B$\sim
10^{-2}-10^{-3}$ dru, $\sigma \sim 5\times 10^{-8}$ pb \\
GENIUS 100 & Ge ioniz 100 kg, $E_{Thr}\sim$10 keV \\
(GENINO) & B$\sim 10^{-3}-10^{-5}$ dru, $\sigma \sim 5\times
10^{-8}-2\times 10^{-9}$ pb
\\
GEDEON & Ge ioniz 28-112 kg, B$\sim 2\times 10^{-3}$ dru ($>$10
keV) $\sigma \sim 10^{-7}-10^{-8}$ pb (40-200 GeV) \\ \hline

 & THE FUTURE~($>$2005-2007)  \\ \hline
DRIFT 10 &   Xe 10 m$^{3}$ TPC, $\sigma \sim 10^{-}8$ pb \\
ZEPLIN-MAX & Xe Two-Phase, $\sigma \sim 10^{-10}$ pb \\
GENIUS & Ge ioniz 1-10 Tons, $\sigma \sim 10^{-9}-10^{-10}$ pb
\\
DRIFT-1 ton & Xe 1 Ton TPC, $\sigma \sim 10^{-10}-10^{-11}$ pb
\\
\hline
\end{tabular}
\end{center}
\end{table*}

\normalsize

To reach the lowest rates predicted ($10^{-5}$ c/kgday) in
SUSY-WIMP-nucleus interaction, or in other words, to explore
coherent interaction cross-sections of the order of
$10^{-9}-10^{-10}$pb, substantial improvements have to be
accomplished in pursuing at its best the strategies reviewed in
this talk, with special emphasis in discriminating the type of
events. These strategies must be focussed in getting a  much lower
background (intrinsic, environmental, ...) by improving
radiopurity and shielding. The nuclear recoil discrimination
efficiency should be optimized going from above 99.7\% up to
99.9\% at the same time that the energy $E_{vis}$ at which
discrimination applies should be lowered. The measurement of the
parameters used to discriminate background from nuclear recoils
should be improved and finally one needs to increase the target
masses and guaranty a superb stability over large exposures. With
these purposes various experiments and a large R+D activity are
under way. Some examples are given in Table \ref{tabla8}. The
conclusion is that the search for WIMPs is well focused and should
be further pursued in the quest for their identification.

\section*{Acknowledgments}

I wish to thank S. Cebri\'{a}n and I.G. Irastorza for their
invaluable collaboration in the making of the exclusion plots and
to J. Morales for useful discussions. The present work was
partially supported by the CICYT and MCyT (Spain) under grant
number AEN99-1033 and by the EU Network contract ERB
FMRX-CT98-0167.


\end{document}